\documentstyle[amsfonts,aps]{revtex}

\begin{document}
\author{Mario Castagnino}
\address{CONICET - Instituto de F\'isica de Rosario\\
Av. Pellegrini 250, Rosario, Argentina}
\author{Olimpia Lombardi}
\address{CONICET - Universidad de Buenos Aires. \\
Pu\'{a}n 470, Buenos Aires, Argentina}
\title{The self-induced approach to decoherence in cosmology}
\maketitle

\begin{abstract}
In this paper we will present the {\it self-induced approach} to
decoherence, which does not require the interaction between the system and
the environment: decoherence in closed quantum systems is possible. This
fact has relevant consequences in cosmology, where the aim is to explain the
emergence of classicality in the universe conceived as a closed
(non-interacting) quantum system. In particular, we will show that the
self-induced approach may be used for describing the evolution of a closed
quantum universe, whose classical behavior arises as a result of decoherence.
\end{abstract}

\section{Introduction}

During the last years, the theory of decoherence has became the new
orthodoxy in the quantum physicists community. At present, decoherence is
studied and tested in many areas such as atomic physics, quantum optics,
condensed matter, etc. (see references in \cite{PZ2000}, \cite{Z2001}).
Following the initial proposal of Zeh \cite{Zeh}, the theory was
systematized and developed in a great number of works. According to Zurek 
\cite{Z1991}, \cite{Z1994}, decoherence is a process resulting from the
interaction between a quantum system and its environment; this process
singles out a preferred set of states, usually called ''pointer basis'',
that determines which observables will receive definite values. This means
that decoherence leads to a sort of selection which precludes all except a
small subset of the states in the Hilbert space from behaving in a classical
manner. Arbitrary superpositions are dismissed, and the preferred states
become the candidate to classical states: they correspond to the definite
readings of the apparatus pointer in quantum measurements, as well to the
points in the phase space of a classical dynamical system. {\it %
Environment-induced-superselection} ({\it einselection}) is a consequence of
decoherence.

In this paper we will present a new approach to decoherence, different from
the mainstream einselection approach. From this new perspective, that we
will call {\it self-induced approach}, decoherence does not require the
interaction between the system and the environment: decoherence in closed
quantum systems is possible. This fact has relevant consequences in
cosmology, where the aim is to explain the emergence of classicality in the
universe conceived as a closed (non-interacting) quantum system. In
particular, we will show that the self-induced approach may be used for
describing the evolution of a closed quantum universe, whose classical
behavior arises as a result of decoherence.

\section{The einselection approach in cosmology}

According to Zurek, einselected states are distinguished by their stability
in spite of the monitoring environment. In Paz and Zurek's words \cite
{PZ2000}, {\bf ''the environment distills the classical essence of a quantum
system''}. This means that, from the einselection view, the split of the
universe into the degrees of freedom which are of direct interest to the
observer (the system) and the remaining degrees of freedom (the environment)
is absolutely essential for decoherence. Such a split is necessary, not only
for explaining quantum measurement, but also for understanding que quantum
origin of the classical world. In fact, the einselection approach considers
the problem of the transition from quantum to classical as the core of the
problem: quantum measurement is conceived as a particular case of the
general phenomenon of the emergence of classicality. In addition, if
classicality only emerges in open quantum systems, it must always be
accompanied by other manifestations of openness, such as dissipation of
energy into the environment. Zurek \cite{Z2001} even considers that the
prejudice which seriously delayed the solution of the problem of the
transition from quantum to classical is itself rooted in the fact that the
role of the openness of a quantum system in the emergence of classicality
was ignored for a very long time.

In summary, decoherence explains the emergence of classicality, but only
open systems can decohere. The question is: what about the universe as a
whole? If, as Zurek himself admits, the universe is, by definition, a closed
system, then it cannot decohere. How to explain, then, the classical
behavior of stars, galaxies and clusters? Zurek \cite{Z1994} considers this
possible criticism: if the universe as a whole is a single entity with no
''outside'' environment, any resolution involving its division into systems
seems unacceptable. Zurek's answer to this objection is based on his
particular conception about the nature of quantum mechanics: for him, the
aim of the theory is to establish the relationships between formal results
and observer's perceptions. And perception is an information processing
function carried out by a physical system, the brain. The brain is conceived
as a massive, neural network-like computer very strongly coupled to its
environment, and the environment plays the role of a commonly accessible
internet-like data base, which allows the observer to make copies of the
records concerning the states of the system with no danger of altering it 
\cite{Z1998}. The stability of the correlations between the state of the
observer's brain and the state of the environment on the one hand, and
between the state of the environment and the state of the observed system on
the other, is responsible for the perception of classicality.

This means that, for the einselection approach, the problem of the
transition from quantum to classical amounts to the question ''why we don't
perceive superpositions?'' \cite{Z1998}. In other words, the task is to
explain, not the emergence of classicality, but {\it our perception of
classicality}. But this position would hardly convince the cosmologist, who
conceives the universe as a single closed object with no other object to
interact with. In the cosmological context, the wave function of the
universe describes, not the system of everything except the observers'
brains, but the universe as a whole. Nevertheless, cosmology tries to
explain, with the universal wave function, the evolution of a closed quantum
universe where the classical behavior described by general relativity
emerges. If we take Zurek's position seriously, without the assumption of a
division of the universe into individual systems the problem of the
emergence of classicality has no solution.

At this point, it could be noted that the einselection approach has been
applied to the cosmological level with interesting results. This is
certainly true, but does no undermine the closed-universe objection. In the
works where the einselection approach is used in cosmology, the general
strategy consists in splitting the universe into some degrees of freedom
which represent the ''system'' of interest, and the remaining degrees of
freedom that are supposed to be non accessible and, therefore, play the role
of an internal environment. For instance, in quantum field theory, it is
usual to perform a decomposition on a scalar field $\phi $, $\phi =\phi
_S+\phi _E$, where $\phi _S$ denotes the system field and $\phi _E$ denotes
the environment field; when it is known that the background field follows a
simple classical behavior, the scalar field is decomposed according to $\phi
=\phi _c+\phi _q$, where the background field $\phi _c$ plays the role of
the system and the fluctuation field $\phi _q$ plays the role of the
environment (see \cite{Calzetta}). This means that, strictly speaking, it is
not the universe what decoheres, but a subsystem of the universe: we
perceive a classical universe because there are unaccessible degrees of
freedom that act as an environment.

These considerations allows us to point out the weakest spot of the
einselection program. When this approach is applied to the universe (and, in
general, to any system with internal environment), the space of the
observables which will behave classically is assumed in advance: the
distinction between the system's degrees of freedom and the environmental
degrees of freedom is established in such a way that the system decoheres in
some observables of that space. This means that the split of the whole must
be decided case by case: there is not a general criterion for discriminating
between system and environment. In fact, in the case of the decomposition of
the scalar field $\phi $ previously mentioned, different criteria are used:
sometimes the decomposition is performed on the basis of the length, mass or
momentum scales of the system and the environment, sometimes the system
field is considered as containing the lower modes of $\phi $ and the
environment as containing the higher modes (see \cite{Calzetta}). Zurek \cite
{Z1998} recognizes that this lack of a general criterion for deciding where
to place the ''cut'' between system and environment is a serious difficulty
of his proposal: {\bf ''In particular, one issue which has been often taken
for granted is looming big, as a foundation of the whole decoherence
program. It is the question of what are the 'systems' which play such a
crucial role in all the discussions of the emergent classicality. This issue
was raised earlier, but the progress to date has been slow at best''}.

As we will see, the self-induced approach to decoherence overcomes these
problems to the extent that it does not require the openness of the system
of interest and its interaction with the environment.

\section{The self-induced approach to decoherence}

This approach relies on the general idea that the interplay between
observables and states is a fundamental element of quantum mechanics (see 
\cite{LC1998a}). The departing point consists in adopting an algebra of
observables ${\cal A}$ as the primitive element of the theory: quantum
states are represented by linear functionals over ${\cal A}$. In the
original formulation of the algebraic formalism, the algebra of observables
is a C*-algebra. The GNS theorem (Gel'fand-Naimark-Segal) proves that the
traditional Hilbert space formalism is a particular representation of this
algebraic formalism; the algebra of observables is thereby given a concrete
representation as a set of self-adjoint bounded operators on a separable
Hilbert space. Nevertheless, it is well known that the C*-algebraic
framework does not admit unbounded operators; therefore, it is necessary to
move to a less restrictive framework in order to accommodate this kind of
operators. The self-induced approach adopts a {\it nuclear algebra} \cite
{Treves} as the algebra of observables ${\cal A}$: its elements are nuclei
or kernels, that is, two variables distributions that can be though of as
generalized matrices \cite{CO2001}. By means of a generalized version of the
GNS theorem \cite{IC1999} it can be proved that this nuclear formalism has a
representation in a rigged Hilbert space: the appropriate rigging provides a
mathematical rigorous foundation to unbounded operators (see \cite{Belanger
and Thomas}). In fact, the nuclear spectral theorem of Gel'fand and Maurin
establishes that, under very general mathematical hypotheses (quite
reasonable from a physical point of view), for every CSCO (complete set of
commuting observables) of essentially self-adjoint unbounded operators,
there is a rigged Hilbert space where such a CSCO can be given a generalized
eigenvalue decomposition, meaning that a continuum of generalized
eigenvalues and eigenvectors may thereby be associated with it. In order to
find the appropriate rigging, the nuclear algebra is used to generate two
additional topologies: one of them corresponds to a nuclear space, which is
the space of generalized observables $V_O$; the other corresponds to the
dual of the space $V_O$, and this is the space $V_{S\text{ }}$of states.

Following \cite{LC1998a}, \cite{Antoniou}, \cite{LC1998b}, we will symbolize
an observable belonging to $V_O$ by a round ket $|O)$ and a state belonging
to $V_S$ by a round bra $(\rho |$. The result of the action of the round bra 
$(\rho |$ on the round ket $|O)$ is the mean value of the observable $|O)$
in the state $(\rho |:$%
\begin{equation}
\langle O\rangle _\rho =(\rho |O)  \label{3.1}
\end{equation}
If the basis is discrete, $\langle O\rangle _\rho $ can be computed as
usual, that is, as $Tr(\rho \,O)$. But if the basis is continuous, $Tr(\rho
\,O)$ is not well defined; nevertheless, $(\rho |O)$ can always be
rigorously defined since $(\rho |$ is a linear functional belonging to $V_{S%
\text{ }}$ acting onto an operator $|O)$ belonging to $V_{O\text{ }}$.

In order to see how decoherence works from the new approach, let us consider
the simplest case, a quantum system whose Hamiltonian has a continuous
spectrum: 
\begin{equation}
H\ |\omega {\bf \rangle }=\omega \ |\omega {\bf \rangle }\qquad \omega \in
\left[ 0,\infty \right)  \label{3.2}
\end{equation}
where $\omega $ and $|\omega {\bf \rangle }$ are the generalized eigenvalues
and eigenvectors of $H$ respectively. The CSCO of this system is just $\{H\}$%
. A generic observable $|O)$ can be expressed in terms of the eigenbasis $%
\left\{ |\omega \rangle \langle \omega ^{\prime }|\right\} $ as: 
\begin{equation}
|O)=\int \int \hat O(\omega ,\omega ^{\prime })\,|\omega \rangle \langle
\omega ^{\prime }|\,d\omega \,d\omega ^{\prime }=\int \int \hat O(\omega
;\omega ^{\prime })\,|\omega ;\omega ^{\prime })\,d\omega \,d\omega ^{\prime
}  \label{3.3}
\end{equation}
where $|\omega ;\omega ^{\prime })=|\omega \rangle \langle \omega ^{\prime
}|\,$ and $\hat O(\omega ,\omega ^{\prime })$ represents the coordinates of
the kernel $|O)$. The Hamiltonian in the eigenbasis $\left\{ |\omega ;\omega
^{\prime })\right\} $ reads: 
\begin{equation}
H=\int \omega \,|\omega \rangle \langle \omega |\,d\omega \,=\int \int
\omega \,\delta (\omega -\omega ^{\prime })\,|\omega ;\omega ^{\prime
})\,d\omega \,d\omega ^{\prime }  \label{3.4}
\end{equation}
Then, $\omega \,\delta (\omega -\omega ^{\prime })$ must be one of the $\hat %
O(\omega ,\omega ^{\prime })$, since $H$ is one of the observables belonging
to $V_{O\text{ }}$. Moreover, all the observables which commute with $H$ and
share the eigenbasis $\left\{ |\omega ;\omega ^{\prime })\right\} $ must
have the following form: 
\begin{equation}
|O)=\int \int O(\omega )\,|\omega \rangle \langle \omega |\,d\omega =\int
\int O(\omega )\,\delta (\omega -\omega ^{\prime })\,\,|\omega ;\omega
^{\prime })\,d\omega \,d\omega ^{\prime }  \label{3.5}
\end{equation}
where $O(\omega )$ supplies the values of the components of $|O)$ in the
basis $\left\{ |\omega ;\omega ^{\prime })\right\} $. Therefore, $O(\omega
)\,\delta (\omega -\omega ^{\prime })$ must be one of the $\hat O(\omega
,\omega ^{\prime })$. But, of course, we need also observables which do not
commute with $H$ and whose $\hat O(\omega ,\omega ^{\prime })$ are different
than $O(\omega )\,\delta (\omega -\omega ^{\prime })$; then, with no loss of
physical generality we can postulate as a general case: 
\begin{equation}
\hat O(\omega ,\omega ^{\prime })=O(\omega )\,\delta (\omega -\omega
^{\prime })+O(\omega ,\omega ^{\prime })  \label{3.6}
\end{equation}
where $O(\omega ,\omega ^{\prime })$ is a regular function whose precise
mathematical properties are listed in \cite{CL2000a}\footnote{%
Since any singular kernel can be approximated by a regular one, the $%
O(\omega )\,\delta (\omega -\omega ^{\prime })+O(\omega ,\omega ^{\prime })$
are dense en the set of the $\hat O(\omega ,\omega ^{\prime })$. Therefore,
we do not loose physical generality, in the sense that the $O(\omega
)\,\delta (\omega -\omega ^{\prime })+O(\omega ,\omega ^{\prime })$ have all
the required physical properties up to any order and, then, they are
experimentally indistinguishable from the $\hat O(\omega ,\omega ^{\prime })$%
.}. Therefore, a generic observable $|O)$ reads (see \cite{van Hove}): 
\begin{equation}
|O)=\int O(\omega )\,\,|\omega )\,d\omega +\int \int O(\omega ,\omega
^{\prime })\,\,|\omega ;\omega ^{\prime })\,d\omega \,d\omega ^{\prime }
\label{3.7}
\end{equation}
where $|\omega )=|\omega \rangle \langle \omega |$ and$\,$ $|\omega ;\omega
^{\prime })=|\omega \rangle \langle \omega ^{\prime }|$ are the generalized
eigenvectors of the observable $|O)$. We will call the first term of the
r.h.s of eq.(7) $O_S$ (the singular part of the observable $|O)$), and the
second term of the r.h.s of eq.(7) $O_R$ (the regular part of the observable 
$|O)$)

The observables $|O)$ of the form (\ref{3.7}) define what we will call {\it %
''van Hove space'', }$V_O^{VH}\subset V_O^{}$; $\left\{ |\omega ),|\omega
;\omega ^{\prime })\right\} $ is the basis of $V_O^{VH}$. On the other hand,
states are represented by linear functionals belonging to a space $V_S^{VH}$%
, which is the dual of $V_O^{VH}$; therefore, a generic state $(\rho |$ can
be expressed: 
\begin{equation}
(\rho |=\int \rho (\omega )\,(\omega \,|\,\,d\omega +\int \int \rho (\omega
,\omega ^{\prime })\,\,(\omega ;\omega ^{\prime }\,|\,d\omega \,d\omega
^{\prime }  \label{3.8}
\end{equation}
where $\rho (\omega ,\omega ^{\prime })$ is a regular function, and $\rho
(\omega )$ and $\rho (\omega ,\omega ^{\prime })$ satisfy the properties $%
\rho \geq 0,$ $(\rho |I)=1$ (where $|I)$ is the identity operator) and those
listed in \cite{CL2000a}. $\left\{ (\omega \,|\,,(\omega ;\omega ^{\prime
}\,|\,\right\} $, the basis of $V_S^{VH}$, is the cobasis of $\left\{
|\omega ),|\omega ;\omega ^{\prime })\right\} $ defined by the following
relations\footnote{%
These are the generalization of the relations between the basis $\left\{
|i\rangle \right\} $ and the cobasis $\left\{ \langle j|\,\right\} $ in the
discrete case: $\langle j|\,i\rangle =\delta _{ij}$.$\,$}:

\begin{equation}
(\omega \,|\omega )=\delta (\omega -\omega ^{\prime })\qquad (\omega ;\omega
^{\prime \prime }\,|\omega ^{\prime };\omega ^{\prime \prime \prime
})=\delta (\omega -\omega ^{\prime \prime })\ \delta (\omega ^{\prime
}-\omega ^{\prime \prime \prime })\qquad (\omega \,|\omega ^{\prime };\omega
^{\prime \prime })=0  \label{3.9}
\end{equation}

Given the expressions (\ref{3.7}) and (\ref{3.8}) for $|O)$ and $(\rho |$
respectively, decoherence follows in a straightforward way. According to the
unitary von Neumann equation, the evolution of $(\rho |$ is given by: 
\begin{equation}
(\rho (t)|=\int \rho (\omega )\,(\omega \,|\,d\omega +\int \int \rho (\omega
,\omega ^{\prime })\,\,e^{-i(\omega -\omega ^{\prime })t}\,(\omega ;\omega
^{\prime }\,|\,d\omega \,d\omega ^{\prime }  \label{3.10}
\end{equation}
Therefore, the mean value of the observable $|O)$ in the state $(\rho (t)|$
reads: 
\begin{equation}
\langle O\rangle _{\rho (t)}=(\rho (t)|O)=\int \rho (\omega )\,O(\omega
)\,d\omega +\int \int \rho (\omega ,\omega ^{\prime })\,\,e^{-i(\omega
-\omega ^{\prime })t}\,O(\omega ,\omega ^{\prime })\,d\omega \,d\omega
^{\prime }  \label{3.11}
\end{equation}
Since $\rho (\omega ,\omega ^{\prime })\,\,$and $O(\omega ,\omega ^{\prime
}) $ are regular functions (see \cite{LC1998a} for details), if we take the
limit for $t\rightarrow \infty $, we can apply the Riemann-Lebesgue theorem,
according to which the second term of the right hand side of the last
equation vanishes. Therefore: 
\begin{equation}
\lim_{t\rightarrow \infty }\,\langle O\rangle _{\rho (t)}=\lim_{t\rightarrow
\infty \,}\,(\rho (t)|O)=\int \rho (\omega )\,O(\omega )\,d\omega
\label{3.12}
\end{equation}
But this integral is equivalent to the mean value of the observable $O$ in a
new state $(\rho _{*}|$: 
\begin{equation}
(\rho _{*}|=\int \rho (\omega )\,\,(\omega \,|\ d\omega  \label{3.13}
\end{equation}
where the off-diagonal terms have vanished. Therefore, we obtain the weak
limit: 
\begin{equation}
\lim_{t\rightarrow \infty }\,\langle O\rangle _{\rho (t)}=\langle O\rangle
_{\rho _{*}}  \label{3.14}
\end{equation}

The next step is to study the formalism under the Wigner transformation $%
"symb"$. Everything behaves in the usual way for the regular parts of $|O)$
and $(\rho |$, since these parts satisfy the hypotheses of papers \cite
{Wigner}. The problem consists in defining the Wigner transformation for the
singular parts. The singular parts of observables and states read:

\begin{equation}
O_{S}=\int O(\omega )\,\,|\omega )\,d\omega =O(H)\qquad \rho _{S}=\int \rho
(\omega )\,(\omega \,|\,d\omega   \label{º2.18}
\end{equation}
Therefore $O_{S}$ is a function of the Hamiltonian: 
\[
H=\int \omega \,\,|\omega )\,d\omega 
\]
Using the well know properties of the Wigner integral we have that: 
\begin{equation}
symbO_{S}=O_{S}^{W}(q,p)=O(H^{W}(q,p))+0(\frac{\hbar ^{2}}{S^{2}})
\label{º2.20}
\end{equation}
where $S$ is the characteristic action of the system. Then, in the
particular case where $O(\omega )=\delta (\omega -\omega ^{\prime })$ we
have from eq. (\ref{º2.20}): 
\begin{equation}
symb|\omega ^{\prime }\rangle \langle \omega ^{\prime }|=\delta
(H^{W}(q,p)-\omega ^{\prime })  \label{º2.21}
\end{equation}
where we have disregarded the $0(\frac{\hbar ^{2}}{S^{2}})$ as we will
always do below. Moreover, for regular functions: 
\begin{equation}
(\rho |O)=(symb\rho |symbO)=\int \rho ^{W}(q,p)\,O^{W}(q,p)\,dq\,dp
\label{norm}
\end{equation}
We will adopt the same equation for the singular parts. This will allow us
to define $symb\rho _{S}$ as satisfying:

\begin{equation}
(symb\rho _S|symbO_S)=(\rho _S|O_S)  \label{º3.9}
\end{equation}
In doing so we are repeating what we have done at the quantum level when we
defined the functional $(\rho |.$

>From eq. (\ref{º2.21}) we know that:

\begin{equation}
symb|\omega ^{\prime })=\delta (H^{W}(q,p)-\omega ^{\prime })  \label{º3.10}
\end{equation}
It is clear that we cannot normalize this function with the variables and in
the domain of integration of (\ref{norm}). In fact, using the canonical
variable $-t$, the canonical variable conjugated to $H,$ for any function
like $symb|\omega ^{\prime },p^{\prime })=\delta (H^{W}(\phi )-\omega
^{\prime })$, which is a constant for $t,$ the integral will turn out to be
infinity. So these functions $f(H)$ are not classical densities since they
do not belong to $L_{1},$ if defined integrating over the whole phase space
,and they must be normalized in a different way. This fact is not surprising
since they are singular functions. But let us observe that, in general:

\begin{equation}
O_S(\phi )=symb\int_0^\infty O(\omega )\,|\omega )\,d\omega =\int_0^\infty
O(\omega )\,\delta (H^W(\phi )-\omega ^{\prime })\,d\omega
\end{equation}
which is a function independent of $-t$, and can be normalized (if
necessary) imposing the following conditions:

i.-We integrate only over the momentum space $H$ (i.e. not over $-t)$,
precisely:

\begin{equation}
||O_{S}(\phi )||=\int dH\int_{0}^{\infty }|O(\omega ,p)|\,\delta (H-\omega
^{\prime })d\omega =\int |O(H)|\,dH
\end{equation}

ii.- We chose the regular function $O(\omega )$ in the space $L_{1}$ of the
momentum, that is to say: 
\begin{equation}
\int |O(\omega )|\,d\omega <\infty
\end{equation}
So, we will normalize all the $f(H)$ (if necessary) in this way, and we will
perform {\it all the integrations in the} ${\cal L}_{S\text{ }}${\it space}
by this method. In particular we will use this way of integration when {\it %
defining functionals of }${\cal L}_{S\text{ }}^{\prime }$

Then, in order to satisfy eq.(\ref{3.9}), necessarily: 
\begin{equation}
\rho _{S\omega }^W(q,p)=symb(\omega ^{\prime }|=\delta (H^W(q,p)-\omega
^{\prime })  \label{º3.19}
\end{equation}
a result already obtained in \cite{CL2000b} (eqs.(34) and (35)) with
different methods. So, $\rho _{S\omega }^W(q,p)$ corresponds to a density
function extremely peaked over the classical trajectory defined by the
conservation law $H^W(q,p)=\omega ^{\prime }$.

>From eq.(\ref{º2.18}): 
\begin{equation}
\rho _S^W(q,p)=\int_0^\infty \rho (\omega )\,\rho _{S\omega
}^W(q,p)\,d\omega =\int_0^\infty \rho (\omega )\,\delta (H^W(q,p)-\omega
^{\prime })\,d\omega =\rho (H^W(q,p))  \label{º3.20}
\end{equation}
and it is a {\it constant of the motion, }as it was expected. Moreover, $%
\rho _S^W(q,p)\geq 0$ since $\rho _S(\omega )\geq 0$. This means that $\rho
_S^W(q,p)$ is the statistical ensemble of the density functions $\rho
_{S\omega }^W(q,p)$ extremely peaked over the classical trajectories defined
by the conservation law $H^W(q,p)=\omega ^{\prime }$ and weighted by the
probabilities $\rho (\omega )$. From eq.(\ref{3.14}) we know that only the
singular part must be considered after decoherence. Thus, for $t\rightarrow
\infty $ the classical density $\rho _S^W(q,p)$ is resolved as a set of
classical trajectories.

At this point, we have defined the Wigner transformation both for the
regular parts and for the singular parts of observables and states. As the
result, the singular parts share the same usual properties with the regular
parts, since we have postulated such properties to define the Wigner
transformation of the singular parts. In fact, these usual properties follow
from eq.(16) (see \cite{CL2000b} for details). E.g., eq.(12) is also valid
when the Wigner transformations are involved: 
\begin{equation}
\lim_{t\rightarrow \infty \,}\,(\rho ^{W}|O^{W})=\int \rho
_{*}^{W}(H)\,O_{S}^{W}(H)\,dH  \label{25'}
\end{equation}
This means that any observable $|O)$ becomes $O_{S}^{W}(q,p)$ in the limit $%
t\rightarrow \infty $, and behaves in a classical way.

After this presentation of the formalism in the simplest case, let us study
the general case. In general, we must consider a CSCO, $\left\{
H,O_{1},...,O_{n}\right\} $, whose eigenvectors are $|\omega
,o_{1},...,o_{n}\rangle $. In this case, $(\rho _{*}|$ will be diagonal in
the variables $\omega ,\omega ^{\prime }$ but not in general in the
remaining variables. Therefore, a further diagonalization of $(\rho _{*}|$
is necessary: as the result, a new set of eigenvectors $\left\{ |\omega
,p_{1},...,p_{n}\rangle \right\} $, corresponding to a new CSCO $\left\{
H,P_{1},...,P_{n}\right\} $ emerges. This set defines the eigenbasis $%
\left\{ |\omega ,p_{1},...,p_{n}),|\omega ,p_{1},...,p_{n};\omega ^{\prime
},p_{1}^{\prime },...,p_{n}^{\prime })\right\} $ of the van Hove space of
observables $V_{O}^{VH}$, where: 
\begin{eqnarray}
|\omega ,p_{1},...,p_{n}) &=&|\omega ,p_{1},...,p_{n}\rangle \,\langle
\omega ,p_{1},...,p_{n}|  \label{3.15} \\
|\omega ,p_{1},...,p_{n};\omega ^{\prime },p_{1}^{\prime },...,p_{n}^{\prime
}) &=&|\omega ,p_{1},...,p_{n}\rangle \,\langle \omega ^{\prime
},p_{1}^{\prime },...,p_{n}^{\prime }|  \nonumber
\end{eqnarray}
$(\rho _{*}|$ will be completely diagonal in the cobasis of states $\left\{
(\omega ,p_{1},...,p_{n}|,(\omega ,p_{1},...,p_{n};\omega ^{\prime
},p_{1}^{\prime },...,p_{n}^{\prime }|\right\} $ corresponding to the new
eigenbasis of $V_{O}^{VH}$ (see \cite{CL2000b} for details). And, most
important, for this system eq.(\ref{º3.20}) reads: 
\begin{equation}
\rho _{S}^{W}(q,p)=\sum_{p}\int_{0}^{\infty }\rho (\omega )\,\rho _{S\omega
p_{1},...,p_{n}}^{W}(q,p)\,d\omega =\sum_{p_{{}}}\int_{0}^{\infty }\rho
(\omega )\,\delta (H^{W}(q,p)-\omega ^{\prime })\,\delta
(P_{i}^{W}(q,p)-p_{i})\,d\omega   \label{3.16}
\end{equation}
where the density $\rho _{S\omega p_{1},...,p_{n}}^{W}$ corresponds to a
density function extremely peaked over the classical trajectory defined by
the conservation laws: 
\begin{equation}
H^{W}(q,p)=\omega ^{\prime },\qquad P_{1}^{W}(q,p)=p_{1},\qquad ...\qquad
P_{n}^{W}(q,p)=p_{n}
\end{equation}
Therefore, for $t\rightarrow \infty $ the classical density $\rho
_{S}^{W}(q,p)$ is resolved again as a set of classical trajectories.

As this presentation shows, decoherence does not require the interaction of
the system of interest with the environment: {\it a single closed quantum
system can decohere}. The diagonalization of the density operator does not
depend on the openness of the system but on the continuous spectrum of the
system's Hamiltonian. This means that the problem of providing a general
criterion for discriminating between system and environment vanishes in the
self-induced approach. This fact leads to an additional advantage of the new
way of conceiving decoherence. As we have seen, in many cases the
einselection approach requires to introduce assumptions about the
observables which will behave classically in order to decide where to place
the boundary between system and environment. The new approach, on the
contrary, provides a mathematically precise definition of the observables
regarding to which the system will decohere. In fact, there are two kinds of
such observables:

a) Observables that commute with the Hamiltonian, which are represented by
the singular kernels $O(\omega )\,\delta (\omega -\omega ^{\prime })$.

b) Observables that do not commute with the Hamiltonian, which are
represented by the kernels $O(\omega )\,\delta (\omega -\omega ^{\prime
})+O(\omega ,\omega ^{\prime })$, where $O(\omega ,\omega ^{\prime })$ is a
regular function. In other words, these observables have a regular part $%
O(\omega ,\omega ^{\prime })$ and a singular part $O(\omega )\,\delta
(\omega -\omega ^{\prime })$ in the eigenbasis defined by the system's
Hamiltonian.

This definition is completely general and does not require to introduce any
prior assumption about the classical behavior of certain observables.

When the phenomenon of decoherence is viewed from this new perspective, it
does not need to be conceived as ''{\bf a justification for the persistent
impression of reality}'' \cite{PZ2000}. Classicality is not a perceptual
result of the correlations between the observed system and the observer's
brain though the environment: the emergence of classicality is a consequence
of the own dynamics of a closed quantum system. In other words, from the
self-induced approach, decoherence is a relevant element for explaining the 
{\it emergence} of classicality, not our {\it perception} of classicality.

\section{Decoherence in a closed universe}

If the transition from quantum to classical does not require the split of
the universe into subsystems as a necessary condition, then decoherence can
take part in the account of how the universe as a whole behaves classically.
In this section we will apply the self-induced approach to a simple
quantum-cosmological model in order to show how classicality arises in this
case.

\subsection{The model}

Let us consider the flat Roberson-Walker universe (\cite{Paz and Sinha}, 
\cite{Cast-Gunzig-Lombardo}) with a metric: 
\begin{equation}
ds^2=a^2(\eta )\ (d\eta ^2-dx^2-dy^2-dz^2)  \label{4.1}
\end{equation}
where $\eta $ is the conformal time and $a$ the scale of the universe. Let
us consider a free neutral scalar field $\Phi $ and let us couple this field
with the metric, with a conformal coupling $(\xi =\frac 16)$. The total
action reads $S=S_g+S_f$ $+S_i$, and the gravitational action is: 
\begin{equation}
S_g=M^2\int d\eta \ [-\frac 12\stackrel{\bullet }{a}^2-\,V(a)]  \label{4.2}
\end{equation}
where $M$ is the Planck mass, $\stackrel{\bullet }{a}=da/d\eta ,$ and the
potential $V$ contains the cosmological constant term and, eventually, the
contribution of some form of classical matter. We suppose that $V$ has a
bounded support $0\leq a\leq a_1$. We expand the field $\Phi $ as: 
\begin{equation}
\Phi (\eta ,{\bf x})=\int_{-\infty }^{+\infty }f_{{\bf k}}(\eta )\,e^{-i{\bf %
k\cdot x}}\,d{\bf k}  \label{4.3}
\end{equation}
where the components of ${\bf k\in }{\Bbb R}$ $^3$ are three continuous
variables.

The Wheeler-De Witt equation for this model reads: 
\begin{equation}
H\,\Psi (a,\Phi )=(h_g+h_f+h_i)\,\Psi (a,\Phi )=0  \label{4.4}
\end{equation}
where: 
\[
h_g=\frac 1{2M^2}\,\partial _a^2+M^2\,V(a) 
\]
\[
h_f=-\frac 12\int (\partial _{{\bf k}}^2-k^2f_{{\bf k}}^2)\ d{\bf k} 
\]

\begin{equation}
h_i=\frac 12m^2a^2\int f_{{\bf k}}^2\ d{\bf k}  \label{4.5}
\end{equation}
with $m$ the mass of the scalar field, ${\bf k}/a$ the linear momentum of
the field, and $\partial _{{\bf k}}$ =$\partial /\partial f_{{\bf k}}.$

We can now go to the semiclassical regime using the WKB method (\cite{Hartle}%
), writing $\Psi (a,\Phi )$ as: 
\begin{equation}
\Psi (a,\Phi )=\exp [iM^{2}S(a)]\,\chi (a,\Phi )  \label{4.6}
\end{equation}
and expanding $S$ and $\chi $ as: 
\begin{equation}
S=S_{0}+M^{-1}S_{1}+...,\qquad \chi =\chi _{0}+M^{-1}\chi _{1}+...
\label{4.7}
\end{equation}
To satisfy eq. (\ref{4.4}) at the order $M^{2}$, the principal Jacobi
function $S(a)$ must satisfy the Hamilton-Jacobi equation: 
\begin{equation}
\left( \frac{dS}{da}\right) ^{2}=2V(a)  \label{4.8}
\end{equation}
We can now define the (semi)classical time as a parameter $\eta =\eta (a)$
such that: 
\begin{equation}
\frac{d}{d\eta }=\frac{dS}{da}\,\frac{d}{da}=\pm \sqrt{2V(a)}\,\frac{d}{da}
\label{4.9}
\end{equation}
The solution of this equation is $a=\pm F(\eta ,C),$ where $C$ is an
arbitrary integration constant. Different values of this constant and of the 
$\pm $ sign give different classical solutions for the geometry.

Then, in the next order of the WKB expansion, $\chi $ satisfy a
Schr\"{o}dinger equation that reads: 
\begin{equation}
i\frac{d\chi }{d\eta }=h(\eta )\chi   \label{4.10}
\end{equation}
where: 
\begin{equation}
h(\eta )=h_{f}+h_{i}(a)  \label{4.11}
\end{equation}
precisely: 
\begin{equation}
h(\eta )=-\frac{1}{2}\int \left[ -\frac{\partial ^{2}}{\partial f_{{\bf k}%
}^{2}}+\Omega _{{\bf k}}^{2}(a)f_{{\bf k}}^{2}\right] d{\bf k}  \label{4.12}
\end{equation}
where: 
\begin{equation}
\Omega _{{\bf k}}^{2}(a)=\Omega _{\varpi
}^{2}(a)=m^{2}a^{2}+k^{2}=m^{2}a^{2}+\varpi   \label{4.13}
\end{equation}
where $\varpi =k^{2}$ and $k=|{\bf k|.}$ So the time dependence of the
Hamiltonian comes from the function $a=a(\eta ).$

Let us now consider a scale of the universe such that $a_{out}\gg a_1$. In
this region the geometry is almost constant. Therefore, we have an adiabatic
final vacuum $|0\rangle $ and adiabatic creation and annihilation operators $%
a_{{\bf k}}^{\dagger }$ and $a_{{\bf k}}.$ Then $h=h(a_{out})$ reads: 
\begin{equation}
h=\int \Omega _\varpi a_{{\bf k}}^{\dagger }a_{{\bf k}}d{\bf k}  \label{4.14}
\end{equation}

We can now consider the Fock space and a basis of vectors: 
\begin{equation}
|{\bf k}_1,{\bf k}_2,...,{\bf k}_n\rangle \cong |\{k{\bf \}\rangle =}a_{{\bf %
k}_1}^{\dagger }a_{{\bf k}_2}^{\dagger }...a_{{\bf k}_n}^{\dagger
}...|0\rangle  \label{4.15}
\end{equation}
where we have called $\{k{\bf \}}$ the set ${\bf k}_1,{\bf k}_2,...,{\bf k}%
_n $, where eventually $n$ goes to infinity. The vectors of this basis are
eigenvectors of $h$: 
\begin{equation}
h|\{k{\bf \}\rangle =}\omega |\{k{\bf \}\rangle }  \label{4.16}
\end{equation}
where: 
\begin{equation}
\omega =\sum_{{\bf k\in \{k\}}}\Omega _\varpi =\sum_{{\bf k\in \{k\}}%
}(m^2a_{out}^2+\varpi )^{\frac 12}  \label{4.17}
\end{equation}
We can now use this energy to label the eigenvector as: 
\begin{equation}
|\{k{\bf \}\rangle =}|\omega ,[{\bf k]\rangle }  \label{4.18}
\end{equation}
where $[{\bf k]}$ is the remaining set of labels necessary to define the
vector unambiguously. $\{|\omega ,[{\bf k]\rangle \}}$ is obviously an
orthonormal basis, so eq. (\ref{4.14}) reads: 
\begin{equation}
h=\int \omega \,|\omega ,[{\bf k]\rangle }\langle \omega ,[{\bf k]|\,}%
d\omega \,d[{\bf k]}  \label{4.19}
\end{equation}

\subsection{Decoherence in energy}

In this case, a generic observable $|O)\in V_{O}^{VH}$ reads (see eq.(\ref
{3.7})): 
\begin{equation}
|O)=\int O(\omega ,[{\bf k],}[{\bf k}^{\prime }{\bf ]})\,\,|\omega ,[{\bf k];%
}[{\bf k}^{\prime }{\bf ]})\,d\omega \,d[{\bf k]\,}d{\bf [k}^{\prime }{\bf %
]\,}+\int \int O(\omega ,[{\bf k],}\omega ^{\prime },[{\bf k}^{\prime }{\bf ]%
})\,\,|\omega ,[{\bf k]};\omega ^{\prime },[{\bf k}^{\prime }{\bf ]}%
)\,d\omega \,d[{\bf k]}\,d\omega ^{\prime }\,d[{\bf k}^{\prime }{\bf ]}
\label{4.20}
\end{equation}
where: 
\[
|\omega ,[{\bf k];}[{\bf k}^{\prime }{\bf ]})=|\omega ,[{\bf k]\rangle }%
\langle \omega ,[{\bf k}^{\prime }{\bf ]|})\qquad |\omega ,[{\bf k]};\omega
^{\prime },[{\bf k}^{\prime }{\bf ]})=|\omega ,[{\bf k]\rangle }\langle
\omega ^{\prime },[{\bf k}^{\prime }{\bf ]|}
\]
and a generic state $(\rho |\in V_{S}^{VH}$ can be expressed as (see eq.(\ref
{3.8})): 
\begin{equation}
(\rho |=\int \rho (\omega ,[{\bf k],}[{\bf k}^{\prime }{\bf ]})\,(\omega ,[%
{\bf k];}[{\bf k}^{\prime }{\bf ]}\,|\,\,d\omega \,d[{\bf k]\,}d[{\bf k}%
^{\prime }{\bf ]}+\int \int \rho (\omega ,[{\bf k],}\omega ^{\prime },[{\bf k%
}^{\prime }{\bf ]})\,\,(\omega ,[{\bf k]};\omega ^{\prime }\,,[{\bf k}%
^{\prime }{\bf ]}|\,d\omega \,d[{\bf k]}\,d\omega ^{\prime }\,d[{\bf k}%
^{\prime }{\bf ]}  \label{4.21}
\end{equation}
where $\left\{ (\omega ,[{\bf k];}[{\bf k}^{\prime }{\bf ]}\,|\,\,,(\omega ,[%
{\bf k]};\omega ^{\prime }\,,[{\bf k}^{\prime }{\bf ]}|\right\} $, the basis
of $V_{S}^{VH}$, is the cobasis of $\left\{ |\omega ,[{\bf k];}[{\bf k}%
^{\prime }{\bf ]}),|\omega ,[{\bf k]};\omega ^{\prime },[{\bf k}^{\prime }%
{\bf ]})\right\} $. Then, the mean value of the observable $|O)$ in the
state $(\rho (t)|$ reads: 
\begin{eqnarray}
\langle O\rangle _{\rho (t)} &=&(\rho (t)|O)=\int \rho (\omega ,[{\bf k];}[%
{\bf k}^{\prime }{\bf ]})\,O(\omega ,[{\bf k];}[{\bf k}^{\prime }{\bf ]}%
)\,d\omega \,d[{\bf k]\,}d[{\bf k}^{\prime }{\bf ]}+ \\
&&+\int \int \rho (\omega ,[{\bf k],}\omega ^{\prime },[{\bf k}^{\prime }%
{\bf ]})\,\,e^{-i(\omega -\omega ^{\prime })t}\,O(\omega ,[{\bf k],}\omega
^{\prime },[{\bf k}^{\prime }{\bf ]})\,d\omega \,d[{\bf k]}\,d\omega
^{\prime }\,d[{\bf k}^{\prime }{\bf ]}  \label{4.22}
\end{eqnarray}
Taking the limit for $t\rightarrow \infty $ and applying the
Riemann-Lebesgue theorem, we obtain: 
\begin{equation}
\lim_{t\rightarrow \infty }\,\langle O\rangle _{\rho (t)}=\lim_{t\rightarrow
\infty \,}\,(\rho (t)|O)=\int \rho (\omega ,[{\bf k];}[{\bf k}^{\prime }{\bf %
]})\,O(\omega ,[{\bf k];}[{\bf k}^{\prime }{\bf ]})\,d\omega \,d[{\bf k]\,}d[%
{\bf k}^{\prime }{\bf ]}  \label{4.23}
\end{equation}
And this integral is equivalent to the mean value of the observable $|O)$ in
a new state $(\rho _{*}|$: 
\begin{equation}
(\rho _{*}|=\int \rho (\omega ,[{\bf k];}[{\bf k}^{\prime }{\bf ]}%
)\,\,(\omega ,[{\bf k];}[{\bf k}^{\prime }{\bf ]}\,|\ d\omega \,d[{\bf k]\,}%
d[{\bf k}^{\prime }{\bf ]}  \label{4.24}
\end{equation}
This new state $(\rho _{*}|$ is the equilibrium time-asymptotic state, which
is diagonal the variables $\omega ,\omega ^{\prime }$ as decoherence in
energy requires.

\subsection{Decoherence in the remaining dynamical variables}

In this case, $(\rho _{*}|$ is diagonal in the variables $\omega ,\omega
^{\prime }$ but not in the remaining variables. This means that a further
diagonalization is necessary.

Let us observe that, if we use polar coordinates for ${\bf k,}$ eq.(\ref{4.3}%
) reads:

\begin{equation}
\Phi (x,n)=\int \sum_{lm}\phi _{klm}\,dk  \label{4.25}
\end{equation}
where:

\begin{equation}
\phi _{klm}=f_{k,l}(\eta ,r)\,Y_m^l(\theta ,\varphi )  \label{4.26}
\end{equation}
where $k$ is a continuous variable, $l=0,1,...,;$ $m=-l,...,l;$ and $Y_m^l$
are spherical harmonic functions. So the indices $k,l,m$ contained in the
symbol ${\bf k}$ are partially discrete and partially continuous.

As $(\rho _{*}^{\dagger }|=(\rho _{*}|$, then $\rho ^{*}(\omega ,[{\bf k],}[%
{\bf k}^{\prime }{\bf ]})=\rho (\omega ,[{\bf k],}[{\bf k}^{\prime }{\bf ]})$
and, therefore, there exists a set of vectors $\{|\omega ,[{\bf l}]\rangle
\} $ such that: 
\begin{equation}
\int \rho (\omega ,[{\bf k],}[{\bf k}^{\prime }{\bf ]})\,|\omega ,[{\bf l}%
]\rangle _{[{\bf k}^{\prime }]}\,d[{\bf k}^{\prime }]=\rho (\omega ,[{\bf l]}%
)\,\,|\omega ,[{\bf l}]\rangle _{[{\bf k]}}  \label{4.27}
\end{equation}
namely, $\{|\omega ,[{\bf l}]\rangle \}$ is the eigenbasis of the operator $%
\rho (\omega ,[{\bf k],}[{\bf k}^{\prime }{\bf ]})$. Then $\rho (\omega ,[%
{\bf l]})$ can be considered as an ordinary diagonal matrix in the discrete
indices $l$ and $m$, and a generalized diagonal matrix in the continuous
index $k$\footnote{{E.g.: We can deal with this generalized matrix by
rigging the space $V_S^{VH}$ and using the Gel'fand-Maurin theorem \cite
{Gorini}; this procedure allows us to define a generalized state eigenbasis
for systems with continuous spectrum. It has been used to diagonalize
Hamiltonians with continuous spectra in \cite{Bohm}, \cite{CGG}, \cite
{Laura1}, etc.}}{.} Under the diagonalization process, eq.(\ref{4.24}) is
written as: 
\begin{equation}
(\rho _{*}|=\int U_{[{\bf k}]}^{\dagger [{\bf l}]}\,\rho (\omega ,[{\bf l],}[%
{\bf l}^{\prime }{\bf ]})\,U_{[{\bf k}^{\prime }]}^{[{\bf l}^{\prime
}]}\,U_{[{\bf k}^{\prime }]}^{\dagger [{\bf l}^{\prime \prime }]}\,(\omega ,[%
{\bf l}^{\prime \prime }{\bf ],[l}^{\prime \prime \prime }]\,|\,U_{[{\bf k}%
]}^{[{\bf l}^{\prime \prime \prime }]}\,d\omega \,d[{\bf k]\,}d[{\bf k}%
^{\prime }{\bf ]\,}d[{\bf l]\,}d[{\bf l}^{\prime }{\bf ]\,}d[{\bf l}^{\prime
\prime }{\bf ]\,}d[{\bf l}^{\prime \prime \prime }{\bf ]}  \label{4.28}
\end{equation}
where $U_{[{\bf k}]}^{\dagger [{\bf l}]}$ is the unitary matrix used to
perform the diagonalization and\footnote{$\delta ([{\bf l]-}[{\bf l}^{\prime
}{\bf ])}$ is a Dirac delta for the continuous indices and a Kronecker delta
for the discrete ones.}: 
\begin{equation}
\rho (\omega ,[{\bf l],}[{\bf l}^{\prime }{\bf ]})=\rho (\omega ,[{\bf l]})\
\delta ([{\bf l]-}[{\bf l}^{\prime }{\bf ])}  \label{4.29}
\end{equation}
Since: 
\begin{equation}
\rho (\omega ,[{\bf l],}[{\bf l]})=\rho (\omega ,[{\bf l]})=\int U_{[{\bf l}%
]}^{[{\bf k}]}\,\rho (\omega ,[{\bf k],}[{\bf k}^{\prime }{\bf ]})\,U_{[{\bf %
l}]}^{\dagger [{\bf k}^{\prime }]}\,d[{\bf k}]\,d[{\bf k}^{\prime }]
\label{4.30}
\end{equation}
we can define: 
\begin{equation}
(\omega ,[{\bf l]}|=(\omega ,[{\bf l],[l}]|=\int U_{[{\bf l}]}^{[{\bf k}%
]}\,(\omega ,[{\bf k],[k}^{\prime }]|\,U_{[{\bf l}]}^{\dagger [{\bf k}%
^{\prime }]\dagger }\,d[{\bf k}]\,d[{\bf k}^{\prime }]  \label{4.31}
\end{equation}
We can repeat the procedure with vectors $(\omega ,\omega ^{\prime },[{\bf %
k],[k}^{\prime }]|$ and obtain vectors $(\omega ,\omega ^{\prime },[{\bf l]|}
$. In this way we obtain a diagonalized cobasis $\{(\omega ,[{\bf l]}%
|,(\omega ,\omega ^{\prime },[{\bf l]}|\}.$ So we can now write the
equilibrium state as: 
\begin{equation}
(\rho _{*}|=\int \rho (\omega ,[{\bf l]})\,(\omega ,[{\bf l}]|\,d\omega \,d[%
{\bf l}]  \label{4.32}
\end{equation}
Since vectors $(\omega ,[{\bf l}]|$ can be considered as diagonal in all the
variables, we have obtained decoherence in all the dynamical variables.

\subsection{Emergence of trajectories}

Let us restore the notation $\{l\}=(\omega ,[{\bf l}]),$ $\{k\}=(\omega ,[%
{\bf k])}$ as in eq.(\ref{4.15}), and let us consider the configuration kets 
$|\{x\}\rangle =|\eta ,[{\bf x}]\rangle .$ Since we are considering the
period when $a\sim a_{out}$, the system with Hamiltonian (\ref{4.14}) is
just a set of infinite oscillators with constants $\Omega _{{\bf k}%
}(a_{out}) $ that represents a scalar field with mass $ma_{out}$. Then, we
are just dealing with a classical set of $N$ particles, with coordinates $[%
{\bf x}]$ and momenta $[{\bf k}]$. Therefore, we can introduce the Wigner
function corresponding to the generalized state $|\{l\})$: 
\begin{equation}
\rho _{\{l\}}^W([{\bf x}],[{\bf k}])=\pi ^{-4N}\int (\{l\}|{\bf x+\lambda
\rangle \langle x-\lambda }|)\,e^{2i[{\bf \lambda ]\bullet [k]}%
}\,d^{4n}\lambda  \label{15.1}
\end{equation}
Using the same reasoning that we have used to obtain eq.(\ref{3.16}) (see
also paper \cite{CL2000b}) $\rho _{\{l\}}^W([{\bf x}],[{\bf k}])$ reads: 
\begin{equation}
\rho _{\{l\}}^W([{\bf x}],[{\bf k}])=\prod_i\delta (L_i^W([{\bf x}],[{\bf k}%
])-l_i)  \label{15.2}
\end{equation}
where $L_i^W([{\bf x}],[{\bf k}])$ is the classical observable obtained from 
$L_i$ (that corresponds to indices ${\bf l)}$ via the Wigner integral
(considering $h=L_0$ and including $0$ among the indices $i$). Now, with the
new notation eq.(\ref{4.32}) reads: 
\begin{equation}
(\rho _{*}|=\int \rho _{\{l\}}\,(\{l\}|\,d\{l\}  \label{15.3}
\end{equation}
Then, if we call: 
\begin{equation}
\rho _{*}^W([{\bf x}],[{\bf k}])=\pi ^{-4N}\int (\rho _{*}|{\bf x+\lambda
\rangle \langle x-\lambda }|)\,e^{2i[{\bf \lambda ]\bullet [k]}%
}\,d^{4n}\lambda  \label{15.4}
\end{equation}
we obtain: 
\begin{equation}
\rho _{*}^W([{\bf x}],[{\bf k}])=\rho _{*}^W(L_0^W([{\bf x}],[{\bf k}%
]),L_1^W([{\bf x}],[{\bf k}]),...)  \label{15.5}
\end{equation}
So finally: 
\begin{equation}
\rho _{*}^W([{\bf x}],[{\bf k}])\sim \int \rho _{\{l\}}\,\rho _{*}^W([{\bf x}%
],[{\bf k}])\,\delta (\{L^W\}-\{l\})\,d\{l\}=\int \rho
_{\{l\}}|\prod_i\delta (L_i^W-l_i)\,d\{l\}  \label{15.7}
\end{equation}
The last equation can be interpreted as follows:

i.- $\delta (\{L^W\}-\{l\})$ is a classical density function, strongly
peaked at certain values of the constants of motion $\{l\},$ corresponding
to a set of trajectories, where the momenta are equal to the eigenvalues $%
l_i $ $(i=0,1,2,...)$.

ii.- $\rho _{\{l\}}$ is the probability to be in one of these sets of
trajectories labelled by $\{l\}$. Precisely: if some initial density matrix
is given, from eq.(\ref{15.3}) it is evident that its diagonal terms $\rho
_{\{l\}}$ are the probabilities to find the density function $\delta
(\{L^W\}-\{l\})$ in the corresponding classical equilibrium density function 
$\rho _{*}^W([{\bf x}],[{\bf k}])$, namely, the probabilities of the
trajectories labelled by $\{l\}=(\omega ,[{\bf l])}$.

iii.- Let ${\bf a}$ be the coordinate classically conjugated to ${\bf l}$
and let be ${\bf a}_0$ the coordinate ${\bf a}$ at time $\eta =0$\footnote{%
We could also add a non-relevant equation, something like $t=\omega \eta
+t_0 $, which would define a choice of our clock's time.}, then we obtain
the classical trajectories: 
\begin{equation}
{\bf a}_i{\bf =l}_i\eta {\bf +a}_{0i}  \label{15.7'}
\end{equation}

iv.- Let us now call $\rho _{\{l\}}=p_{\{l\}[{\bf a}_0]}$. Actually, $%
p_{\{l\}[{\bf a}_0]}$ is not a function of ${\bf a}_0$; it is simply a
constant in ${\bf a}_0$, since ${\bf a}_0$ is only an arbitrary point and
our model is spatially homogenous. Then we can write: 
\begin{equation}
p_{\{l\}[{\bf a}_0]}=\int p_{\{l\}[{\bf a}_0]}\prod_{i=1}^n\delta ({\bf a}_i-%
{\bf a}_{0i})\,d[{\bf a}_0]  \label{15.8'}
\end{equation}
In this way we have changed the role of ${\bf a}_0$: it was a fixed (but
arbitrary) point, and now it is a variable that moves all over the space.
Then eq.(\ref{15.7}) reads: 
\begin{equation}
\rho _{*}^W([{\bf x}],[{\bf k}])\sim \int p_{\{l\}[{\bf a}_0{\bf ]}%
}\prod_i^n\delta (L_i^W-l_i)\prod_{j=1}^n\delta ({\bf a}_j-{\bf a}_{0j})\,d[%
{\bf a}_0]\,d\{l\}  \label{15.8'''}
\end{equation}
So, if we call : 
\begin{equation}
\rho _{\{l\}[{\bf a}_0]}^W([{\bf x}],[{\bf k}])=\prod_{i=0}^n\delta
(L_i^W-l_i)\prod_{j=1}^n\delta ({\bf a}_j-{\bf a}_{0j})  \label{15.8''''}
\end{equation}
we have: 
\begin{equation}
\rho _{*}^W([{\bf x}],[{\bf k}])\sim \int p_{\{l\}[{\bf a}_0{\bf ]}}\,\rho
_{\{l\}[{\bf a}_0]}^W([{\bf x}],[{\bf k}])\,d[{\bf a}_0{\bf ]\,}d\{l\}
\label{15.8'''''}
\end{equation}
>From eq.(\ref{15.8''''}) we see that $\rho _{\{l\}[{\bf a}_0]}^W([{\bf x}],[%
{\bf k}])\neq 0$ only in a narrow strip around the classical trajectory (\ref
{15.7'}) defined by the momenta $\{l\}$ and passing through the point $%
\left[ {\bf a}_0\right] $ (actually the density function is as peaked as it
is allowed by the uncertainty principle; its width is essentially a $0(\frac %
\hbar S)$, since the $\delta -$functions of all the equations become Dirac's
deltas only when $\hbar \rightarrow 0$). Therefore, eq.(\ref{15.8'''''})
describes the classical behavior of our model of universe\footnote{%
In this section, as in section IIB, we have faced the following problem: $%
\rho _{*}^W([{\bf x}],[{\bf k}])$ is a ${\bf a}$ constant that we want to
decompose in functions $\rho _{\{l\}[{\bf a}_0]}^W([{\bf x}],[{\bf k}])$
which are different from zero only around the trajectory (\ref{15.7'}) and
therefore are variables in ${\bf a}$. Then, essentially we use the fact that
if $f(x,y)=g(y)$ is a constant function in $x$, we can decompose it as: 
\[
g(y)=\int g(y)\delta (x-x_0)dx_0 
\]
namely, the densities $\delta (x-x_0)$ are peaked in the trajectories $%
x=x_0=const.$, $y=var.$ and, therefore, are functions of $x.$ This
trajectories play the role of those of eq.(\ref{15.8'}).
\par
As all the physics, including the correlations, is already contained in eq.(%
\ref{15.7}), the reader may just consider the final part of this section,
from eq.(\ref{15.8'}) to eq.(\ref{15.8'''''}) a didactical presentation.}.

Let us sum up the main steps of our argument. When $\eta \rightarrow \infty $%
, the quantum density $\rho $ becomes a diagonal density matrix $\rho _{*}$.
The corresponding classical distribution $\rho _{*}^{W}([{\bf x}],[{\bf k}])$
can be expanded as a sum of classical trajectories density functions $\rho
_{\{l\}[{\bf a}_{0}]}^{W}([{\bf x}],[{\bf k}])$, each one weighted by its
corresponding probability $p_{\{l\}[{\bf a}_{0}{\bf ]}}$. Going back to eq.(%
\ref{15.8'''''}), $\rho _{\{l\}[{\bf a}_{0}]}^{W}([{\bf x}],[{\bf k}])=\rho
_{\{{\bf l}_{1},{\bf l}_{2},\ldots {\bf l}_{n}\}[{\bf a}_{0}]}^{W}([{\bf x}%
],[{\bf k}])$ is the density corresponding to the set of $n$ points (let us
say, galaxies), each one of them moving over a trajectory defined by eq.(\ref
{15.7'}), where eventually $n$ goes to infinity. So, as the limit of our
quantum model we have obtained a classical statistical mechanical model, and
the classical realm appears.

In summary, we have proved that the density operator is translated into a
classical density, via a Wigner function, and it is decomposed as a sum of
densities peaked around all possible classical trajectories, each one of
these densities weighted by their own probability. Therefore, our quantum
density operator behaves in its classical limit as a statistical
distribution among a set of classical trajectories. Similar results are
obtained in papers \cite{Zouppas} and \cite{Polarsky}.

\section{Operators and fields}

Up to this point, all our reasoning was made in the context of the
Schr\"{o}dinger picture, as it is usual in quantum mechanics. But in quantum
field theory, on the contrary, the Heisenberg picture is the usual scenario.
However, it is quite easy to reformulate the argument in the new picture,
since our main equations are (14) and (26) for the general case, and (53)
and its corresponding classical version for the cosmological case. In both
cases only the singular parts $O_{S}$ and $\rho _{S}=\rho _{*}$ appear, and
both functions are time-independent. Then, whereas in the Schr\"{o}dinger
picture we have that $O=O(t_{0})$ is time-constant and $\rho (t)\rightarrow
\rho _{*}=\rho _{S}$ is time-constant in the limit, in the Heisenberg
picture $\rho =\rho (t_{0})$ is time-constant and $O(t)\rightarrow
O_{*}=O_{S}$ is time-constant in the limit; nevertheless, both results
coincide since, when $t\rightarrow \infty $, $\rho _{*}=\rho _{S}$ and $%
O_{*}=O_{S}$. Then, if we consider the Heisenberg picture, when $%
t\rightarrow \infty $ the operators $f_{{\bf k}}(\eta )$ of eq.(32) become
singular and also equal to the constants $f_{{\bf k}}$; therefore, the
corresponding $f_{{\bf k}}^{W}$ are constant and classical. The same can be
said for the field $\Phi (\eta ,{\bf x}).$ of eq.(32), that becomes a time
constant $\Phi ({\bf x}).$

With this procedure, in certain sense we have lost the dynamics of the
field. However, it can be recovered if we compute a ''master equation'' from
any asymptotic expansion of the Riemann-Lebesgue theorem or if we make an
analytic continuation in the Liouville complex energy plane as in paper \cite
{Arbo}. Such a master equation coincide with the usual one and also with the
Lindblad approach.

\section{Conclusion}

In this paper our aim was to rigorously present the self-induced approach to
decoherence, according to which the phenomenon of decoherence is the result
of the own dynamics of a closed system governed by a Hamiltonian with
continuous spectrum. From this approach, the interaction between the system
and the environment is not needed and, therefore, a single closed quantum
system can decohere. This feature makes this approach particularly
appropriate for addressing problems in quantum cosmology, since the universe
is, by definition, a quantum closed system with no environment to interact
with. In this context, we have applied the self-induced approach to a
quantum cosmological model, showing that decoherence in energy and in the
remaining dynamical variables obtains with no reference to an environment
and without assuming in advance which observables will behave classically.

\end{document}